
\input amstex
\documentstyle{amsppt}
\overfullrule0pt
\def\fmn#1#2{{F_{#1,#2}}}
\leftline{ hep-th/9507019 }
\vskip1pc
\topmatter
\title  A formula and some operators
\endtitle
\author Seung Hwan Son
\endauthor
\abstract  A formula for some dispersionless equations
and a brief review of the operators which have been used
for the dispersionless KP hierarchy are presented.
\endabstract
\endtopmatter
\document
\head 1. Introduction
\endhead
\footnotetext""{1991 {\it Mathematics Subject Classification.}
Primary 35Q53, 35Q51.}
\footnotetext""{{\it Key words and phrases.}
Dispersionless limit, KdV, KP, Boussinesq, hierarchies, operator.}
\noindent
After the solitonic waves were found by J. Scott Russell in 1843,
a lot of equations describing these kinds of physical phenomena
have been found. And
various hierarchies have been developed from these equations.
By eliminating the dispersive terms, a
lot of dispersionless equations are derived.

To derive  hierarchies, several operators have been invented
and developed. These operators have their own purpose and
advantage.

This article reviews some definitions which have been used
in recent development [13,14,20,22] and some relations
(e.g. equations or identities).
The operators in this article have been used to define and
develop the theories for the dispersionless KP hierarchy.

A formula was found in 1994 by the author and it provides
several equations related to some hierarchies. By the formula,
EQUATION($p,q$) is defined.${}^*$
\footnotetext"* (Caution)"{
\item{} This notation is used only for technical
simplicity of expression since the formula was found
recently [19].
This is another way of using equation
numbers. Therefore, the author
strongly recommends careful use of the notation until
the formula is well-known. (Of course, given notation
has no meaning elsewhere like other usual equation numbers.)}
This formula has an algebraic expression and one can
find an exact expression using this formula for each
positive integers $p$ and $q$.

\head 2. EQUATION($p,q$)
\endhead
Let us use $F_{m,n}$ instead of $\displaystyle\frac{\partial^2}
{\partial t_m\partial t_n}\Bigl(
F(t_1,\ldots,t_r,\ldots)\Bigr)$.
\vskip10pt
\noindent{\bf Definition 2.1.} \qquad EQUATION($p,q$):
\hfill\break
${\displaystyle\sum_{0<i_1<\cdots<i_{k_p}\atop (i_1+1)n_{i_1}
+\cdots+(i_{k_p}+1)n_{i_{k_p}}=p}}
\left(\Bigl(\sum\limits_{j=1}^{k_p} n_{i_j}-1\Bigr)!
\prod\limits_{j=1}^{k_p}{\displaystyle
 \frac{(-\,F_{1,i_j})^{n_{i_j}}}
{n_{i_j}!}}\right)$\hfill\break
\strut \hskip14pc
$+{\displaystyle\sum_{m+n=p} \frac{F_{m,n}}{mn}}=0.$
\hfill\break
where the terms having $\frac{\partial}{\partial t_{q}},\ldots,
\frac{\partial}{\partial t_{kq}},\ldots$ vanish.

\proclaim{Examples 2.2} \rm A table  of these equations
(up to $p=10$) is in [19].
\endproclaim
{\settabs 20\columns
\+& (4,$\infty$)&& $\frac12\fmn11^2-\frac13\fmn13
+\frac14\fmn22=0,$\cr
\+& (4,2)&& $\frac12\fmn11^2-\frac13\fmn13=0$,\cr
\+& (4,3)&& $\fmn11^2+\frac12\fmn22=0$,\cr
\+& (5,$\infty$)&& $\fmn11\fmn12-\frac12\fmn14
+\frac13\fmn23=0,$\cr
\+& (5,3)&& $\fmn11\fmn12-\frac12\fmn14=0,$\cr
\+& (5,4)&& $\fmn11\fmn12+\frac13\fmn23=0.$\cr}
\vskip1pc

\proclaim{Definition 2.3}\rm\hfill\break
\strut\hskip3pc
EQUATION($\cdot,q$)=
$\left\{\,\hbox{\rm EQUATION}(p,q)\,
\bigm| \, p\geq 4\right\}$.\hfill\break
\strut\hskip3pc
EQUATION($p,\cdot$)=
$\left\{\,\hbox{\rm EQUATION}(p,q)\,
\bigm| \, q>1\right\}$.
\endproclaim

\proclaim{Examples 2.4}\rm\qquad\vskip5pt\noindent
$\eqalign{\hbox{\rm EQUATION}(\cdot,\infty)
=&\{\frac12\fmn11^2-\frac13\fmn13+\frac14\fmn22=0,
\fmn11\fmn12-\frac12\fmn14 +\frac13\fmn23=0,\ldots\}\cr
\sim& \hbox{\rm the dispersionless KP hierarchy}\cr}$
\vskip5pt\noindent
$\eqalign{\hbox{\rm EQUATION}(\cdot,2)
=&\{\fmn11^2-\frac13\fmn13=0,
\frac13\fmn11^3-\fmn11\fmn13+\frac35\fmn15-\frac19\fmn33=0,
\ldots\}\cr
\sim& \hbox{\rm the dispersionless KdV hierarchy}\cr}$
\vskip5pt\noindent
$\eqalign{\hbox{\rm EQUATION}(\cdot,3)
=&\{\fmn11^2+\frac12\fmn22=0,\fmn11\fmn12-\frac12\fmn14=0,
\ldots\}\cr
\sim& \hbox{\rm the dispersionless Boussinesq hierarchy}\cr}$
\hfill\break\strut\hskip1pc ...\hfill\break\noindent
$\eqalign{\hbox{\rm EQUATION}(4,\cdot)
=&\{\fmn11^2-\frac13\fmn13=0,
\fmn11^2+\frac12\fmn22=0,
\frac12\fmn11^2-\frac13\fmn13
+\frac14\fmn22=0\}\cr
\sim&\{
u_t-\frac32uu_x=0, (uu_x)_x+\frac12u_{yy}=0,
(u_t-\frac32uu_x)_x-\frac34u_{yy}=0\}.\cr}$
\endproclaim


\head 3. Operators for the dispersionless KP hierarchy
\endhead
\noindent
The following definitions and the theorems  are in
[13,14,20,22].
\proclaim{Definition 3.1}\rm\qquad
($\Omega$ operator):\vskip5pt
\strut \hskip5pc$\Omega=\xi+\sum\limits_{n=1}^\infty
u_{n+1}\xi^{-n}.$
\endproclaim

\proclaim{Definition 3.2}\rm\qquad
$\biggl(\,\, [\cdot]_{\geq s}\quad ;\quad
[\cdot]_{<t}\,\,\biggr)$:
\hfill\break
For integers $s$ and $t$,\vskip5pt
\strut \hskip 5pc
$\left[\sum\limits_{i=-\infty}^{\infty} c_i\xi^i
\right]_{\geq s}$\quad denotes
\quad $\,\,\,\,\sum\limits_{i=s}^{\infty}
c_i\xi^i$\hskip 3pc and
\vskip5pt
\strut\hskip5pc
$\left[\sum\limits_{i=-\infty}^{\infty} c_i\xi^i
\right]_{<t}$\quad denotes
\quad $\sum\limits_{i=-t+1}^{\infty} c_{-i}\xi^{-i}$.
\endproclaim

\proclaim{Examples 3.3}\rm\vskip5pt
$\,[\Omega^2]_{\geq 0}=[(\xi+u_2\xi^{-1}+\cdots)^2]_{\geq 0}
=[\xi^2+2u_2+\cdots]_{\geq0}=\xi^2+2u_2$. And \vskip5pt
$\eqalign{[\Omega^3]_{\geq 0}
=&[(\xi+u_2\xi^{-1}+u_3\xi^{-2}+\cdots)^3]_{\geq 0}\cr
=&[\xi^3+3u_2\xi+3u_3+\cdots]_{\geq 0}\cr
=&\xi^3+3u_2\xi+3u_3.\cr}$
\endproclaim

\proclaim{Remark 3.4}\rm
The dispersionless KP  hierarchy has  the following
Lax representation
of variables  $(x_1,x_2,\ldots)$, ($x_1=x$)
$${ \partial \Omega\over \partial x_n}
=\{ [\Omega^n]_{\geq 0},\Omega\},
\qquad n=1,2,\ldots, \eqno(5.1)$$
where
\def\calb{{\Cal B}}
\def\calll{{\Cal L}}
{\settabs 20\columns
\+&&& $\Omega$  &&:& a series in $\xi$ of the form
$\Omega=\xi+\sum\limits_{n=1}^\infty u_{n+1}
(x_1,x_2,\ldots)\xi^{-n}$,\cr
\+&&&&&& for arbitrary functions $\{u_2(x_1,x_2,\ldots),
u_3(x_1,x_2,\ldots),\ldots\}$,\cr
\+&&& $\{{\Cal B},{\Cal L}\}$ &&:& the Poisson bracket
defined by\cr}
\strut\hskip 8pc
$\displaystyle \{ \calb(\xi,x),\calll(\xi,x)\}
= \frac{\partial \calb(\xi,x)}{\partial \xi}
\frac{\partial \calll(\xi,x)}{\partial x}
-\frac{\partial \calll(\xi, x)}{\partial \xi}
\frac{\partial \calb(\xi,x)}{\partial x}.$
\endproclaim

\proclaim{Remark 3.5} \rm
Let us use $\tau_{dKP}$
to represent the tau function of the dispersionless KP
hierarchy.
\endproclaim

\proclaim{Definition 3.6} \rm\qquad ($\Phi$ operator):
\hfill\break
$$\Phi=\sum_{n=1}^\infty n x_n \Omega^{n-1}
+\sum_{k=1}^\infty v_{k+1}\Omega^{-k-1}.$$
\endproclaim

There are some relations between the operators.
\proclaim{Theorem 3.7}
$\Omega$ and   $\Phi$ have the following relation.
$$[\Omega^n]_{\geq 0}=\Omega^n-\sum_{k=1}^\infty\frac1k\,
\frac{\partial v_{k+1}}{\partial x_n}\Omega^{-k}.$$
\endproclaim

\proclaim{Remark 3.8}\rm
$\tau_{dKP}$ is defined via
$$\frac{\partial \log \tau_{dKP}}{\partial x_n}
=v_{n+1}.$$
\endproclaim

If we substitute $\frac{\partial \log
\tau_{dKP}}{\partial x_n}$
for $v_{n+1}$, we get the following.
\proclaim{Theorem 3.9} If $F=\tau_{dKP}$, then
$$[\Omega^n]_{\geq 0}=\Omega^n-\sum_{k=1}^\infty\frac1k\,
\frac{\partial^2 F}{\partial x_n\partial x_k}\Omega^{-k}.$$
\endproclaim

\head 4. Discussion
\endhead
\noindent
 It is known that the EQUATION($\cdot,\infty$)
is a subset of the dispersionless KP hierarchy.
EQUATION($\cdot,2$) can be regarded
as a subset of the dispersionless KdV hierarchy and
EQUATION($\cdot,3$) can
be regarded as a subset of the dispersionless  Boussinesq
hierarchy.
EQUATION($\cdot,q$) can be
regarded as a new hierarchy for each $q>3$.

Since one can get a specific equation for each $(p,q)$,
several theories related to the solutions of some
dispersionless hierarchies can be used
[8,9,10,13,20].

\vskip2pc
\head Table.\qquad Equation($p,q$).
\endhead
\vskip1pc
\def\frc{\frac}
\def\qqqq#1#2{(#1,#2)\qquad}
\def\nnnn{\vskip8pt}

\qqqq4{$\infty$}
$\frc12\fmn11^2-\frc13\fmn13+\frc14\fmn22=0$\nnnn
\qqqq5{$\infty$}
$\fmn11\fmn12-\frc12\fmn14+\frc13\fmn23=0$\nnnn
\qqqq6{$\infty$}
$\frc13\fmn11^3-\frc12\fmn12^2-\fmn11\fmn13
+\frc35\fmn15-\frc19\fmn33-\frc14\fmn24=0$\nnnn
\qqqq7{$\infty$}
$\fmn11^2\fmn12-\fmn12\fmn13-\fmn11\fmn14+\frc23\fmn16
-\frc16\fmn34-\frc15\fmn25=0$
\vskip2pc

\head References
\endhead
\roster
\item"[1]"
M. J. Ablowitz and H. Segur,
{\it Solitons and the Inverse Scattering Transform}, SIAM (1981).
\item"[2]"
S. Aoyama and Y. Kodama,
{\it A generalized Sato equation and the $W_\infty$ algebra},
Phys. Lett. B {\bf 278} (1992), 56--62.
\item"[3]"
S. Aoyama and Y. Kodama,
{\it The $M$-truncated KP hierarchy and matrix models},
Phys. Lett. B {\bf 295} (1992), 190--198.
\item"[4]"
S. Aoyama and Y. Kodama,
{\it Topological Conformal Field Theory with a Rational
$W$ Potential and the dispersionless KP hierarchy},
Mod. Phy. Lett. A, {\bf 9} No. 27 (1994), 2481--2492.
\item"[5]"
R. Dijkgraaf, H. Verlinde and E. Verlinde,
{\it Loop Equations and Virasoro Constraints in Non-Perturbative
2d Quantum Gravity}, Nucl. Phys. {\bf B348} (1991), 435--456.
\item"[6]"
R. Dijkgraaf, H. Verlinde and E. Verlinde,
{\it Topological Strings in $d<1$}, Nucl. Phys. {\bf B352} (1991),
59--86.
\item"[7]"
J. Gibbons and Y. Kodama,
{\it Solving dispersionless Lax Equations},
Singular Limits of Dispersive Waves, (Plenum, 1994), 61--66.
\item"[8]"
Y. Kodama and J. Gibbons,
{\it A Method for Solving the dispersionless KP hierarchy
and its Exact Solutions. II},
Phys. Lett. A {\bf 135} No. 3 (1989), 167--170.
\item"[9]"
Y. Kodama and J. Gibbons,
{\it Integrability of the dispersionless KP hierarchy},
Proc. of the Workshop on Nonlinear Processes in Physics
(World Scientific, 1990), p. 166.
\item"[10]"
Y. Kodama,
{\it Exact Solutions of Hyperdynamic type Equations having
Infinitely Many Conserved Densities},
Phys. Lett. A {\bf 135} No. 3 (1989), 171--174.
\item"[11]"
B. G. Konopelchenko,
{\it Introduction to Multidimensional Integrable Equations},
Plenum Press, (1992).
\item"[12]"
I. M. Krichever,
{\it Method of Averaging for Two-dimensional Integrable Equations},
Funct. Anal. Appl. {\bf 22} (1988), 200--213.
\item"[13]"
I. M. Krichever,
{\it The dispersionless Lax Equations and Topological Minimal Models},
Commun. Math. Phys. {\bf 143} (1991), 415--426.
\item"[14]"
I. M. Krichever,
{\it The $\tau$-Function of the Universal Whitham Hierarchy,
Matrix Models and Topological Field Theorie}, hep-th/9205110.
\item"[15]"
Y. Ohta, J. Satsuma, D. Takahashi and T. Tokihiro,
{\it An Elementary Introduction to Sato Theory}, Prog. Theoret.
Phys. Supp., {\bf 94} (1988), 210.
\item"[16]"
S. Oishi, {\it A Method of Analysing Soliton Equations by
Bilinearization,} J. Phys. Soc. Jpn, {\bf 48} (1980), 639.
\item"[17]"
M. Sato and Y. Sato,
{\it Soliton Equations as Dynamical Systems on Infinite Dimensional
Grassmann Manifold}, Proc. the U.S.-Japan Seminar, Tokyo, 1982.
\item"[18]"
M. Sato and M. Noumi,
{\it Soliton Equations and the Universal Grassmann Manifolds},
Sophia Univ. Kokyuroku in Math., {\bf 18} (1984).
\item"[19]"
S. H. Son
{\it The equations of some dispersionless limit},
hep-th/9506172.
\item"[20]"
K. Takasaki and T. Takebe,
{\it SDiff(2) KP Hierarchy},
Int. J. Mod. Phys. {\bf A7, Suppl. 1B} (1992), 889--922.
\item"[21]"
K. Takasaki and T. Takebe,
{\it Quasi-Classical Limit of KP Hierarchy, W-Symmetries and
Free Fermions}, Kyoto preprint KUCP-0050/92 (July, 1992).
\item"[22]"
K. Takasaki and T. Takebe,
{\it Integrable hierarchies and dispersionless limit},
hep-th/9405096.
\item"[23]"
E. Witten,
{\it Ground Ring of Two Dimensional String Theory},
Nucl. Phys. {\bf B373} (1992), 187--213.
\endroster
\vskip2pc
{\obeylines
Seung H. Son
Department of Mathematics
University of Illinois at Urbana-Champaign
{\it E-mail address}:  {\tt son\@math.uiuc.edu}
}
\enddocument
\bye